\documentclass[aps,11pt,notitlepage,oneside,onecolumn,nobibnotes,nofootinbib,preprintnumbers,superscriptaddress,showpacs,showkeys]{revtex4}
\usepackage{graphicx}
\graphicspath{{figures/}}

\def\kaos{{\sc Kaos}\@}

\newcommand{\D}{\mathrm{d}}  
  
\begin{document}
% -------------------------------------------------------------------
%
\title{Strangeness physics at MAMI: First results and perspectives}
\author{Patrick Achenbach\\ for the A1 Collaboration}
\email{patrick@kph.uni-mainz.de}
\affiliation{Institut f\"ur Kernphysik, Johannes
  Gutenberg-Universit\"at, Mainz, Germany}

\date{\today}
\begin{abstract}
During the last two years several experimental approaches to strange
systems have been realized at the spectrometer facility of the Mainz
Microtron MAMI. An instrument of central importance for the
strangeness electro-production program is the magnetic spectrometer
\kaos\ that was installed during 2003--8 and is now operated by the A1
collaboration in $(e,e'K)$ reactions on the proton or light nuclei.
Since 2008 kaon production at low four-momentum transfers off a liquid
hydrogen target was studied. The measurements were sensitive to
details of the phenomenological models describing the reaction. Two
very prominent isobar models, Kaon-Maid and Saclay-Lyon A, differ in
the number of contributing nucleon resonances and their longitudinal
couplings at the kinematics measured at MAMI.  In order to use
\kaos\ as a zero-degree double-arm spectrometer a magnetic chicane
comprising two compensating sector magnets was constructed and a new
electron-arm focal-plane detector system was built.
\end{abstract}

 % Baryon production  

% -------------------------------------------------------------------
\maketitle
% -------------------------------------------------------------------

%
% -------------------------------------------------------------------
\section{Introduction}
% -------------------------------------------------------------------
The Mainz Mikrotron MAMI at the Institut f\"ur Kernphysik in Mainz is
an accelerator to study the hadron structure with the electromagnetic
probe.  The machine has been upgraded to 1.5\,GeV electron beam energy
by a harmonic double-sided microtron~\cite{Kaiser2008}.  This fourth
stage, called MAMI-C, was completed in 2007.  With MAMI-C the
threshold beam energy for associated strangeness production off
protons was reached for the first time at MAMI.

The electromagnetic production of kaons off the nucleon provides an
important tool for understanding the dynamics of hyperon-nucleon
systems. Theoretically, the process is described often by effective
Lagrangian models, commonly referred to as ``isobar'' approach. In the
isobaric models the reaction amplitudes are constructed from
lowest-order (so-called tree-level) Born terms with the addition of
extended Born terms for intermediate particles, $N$, $K$, or $Y$
resonances, exchanged in the $s$-, $t$-, and $u$-channels.  A complete
description of the reaction process would require all possible
channels that could couple to the initial and final state. Most of the
model calculations for kaon electrophoto-production have been
performed in the framework of tree-level isobar
models~\cite{Adelseck1990,Williams1992,Bennhold1999}, however, few
coupled-channels calculations exist~\cite{Feuster1999}.  The drawback
of the isobaric models is the large and unknown number of exchanged
hadrons that can contribute in the intermediate state of the
reaction. Depending on which set of resonances is included, very
different conclusions about the strengths of the contributing diagrams
for resonant baryon formation and kaon exchange may be reached. The
Kaon-Maid and Saclay-Lyon models characterise our present
understanding of kaon photoproduction reactions at photon energies
below 1.5\,GeV.  The model SLA is a simplified version of the full
Saclay-Lyon model in which a nucleon resonance with spin-5$/$2 appears
in addition.

The electro-production of strangeness introduces two additional
contributions, that are vanishing for the kinematic point at $Q^2=$
0: the longitudinal coupling of the photons in the initial state, and
the electromagnetic and hadronic form factors of the exchanged
particles. Concerning the differential cross-section the model
variations are strongest at small kaon centre-of-mass angle.  To
conclude, it is fair to say that new experimental data on strangeness
production will challenge and improve our understanding of the strong
interaction in the low energy regime of QCD.

% -------------------------------------------------------------------
\section{Elementary Kaon Electroproduction at MAMI}
% -------------------------------------------------------------------

%
\begin{table}
  \caption{Experimental setting of the kaon electro-production
    beam-time in 2009. The beam energy was 1.508\,MeV, beam intensity
    1--4\,$\mu$A, beam raster size $\pm$ 1\,mm, and the $\ell$H$_2$
    target length 48\,mm.}
  \begin{center}
    \begin{tabular}{ccccccccc} 
      \toprule
      \multicolumn{4}{c}{virt.\ photon + target} 
      & \multicolumn{2}{c}{electron arm}
      & \multicolumn{3}{c}{kaon arm} \\
      \hline
      $\langle Q^2 \rangle$ & $\langle W \rangle$ 
       & $\langle \epsilon \rangle$ & $\langle \omega \rangle$
       & $\langle q^{lab}_{e'} \rangle$ & $\langle \theta^{lab}_{e'} \rangle$
       & $\langle p^{lab}_K \rangle_\Lambda$ & $\langle p^{lab}_K \rangle_\Sigma$
       & $\langle \theta^{lab}_K \rangle$ \\
      (GeV$\!/c$)$^2$ & GeV & (trans.) & GeV
       & GeV$\!/c$ & deg & GeV$\!/c$ & GeV$\!/c$ & deg \\
      \hline
      0.036 & 1.750 & 0.395 & 1.182 & 0.318 & 15.5 
       & 0.642 &  0.466 & $-$31.5 \\
    \hline
    \end{tabular}
  \end{center}
  \label{tab:kinematics}
\end{table}

First experiments on the electro-production of kaons off a liquid
hydrogen target were performed since 2008. Positive kaons were
detected in the \kaos\ spectrometer in coincidence with the scattered
electron into spectrometer~B.  The central spectrometer angle of the
kaon arm was 31.50$^\circ$ with a large angular acceptance in the
dispersive plane of $\vartheta_K=$ 21--43$^\circ$. The spectrometer
setting for electron detection was fixed at the minimum in-plane angle
of $\vartheta_{e'}\approx$ 15$^\circ$, thereby maximising the virtual
photon flux.  The invariant momentum transfer, $\langle Q^2\rangle =
-\langle q^2\rangle$, was very low, the virtual photon energy,
$\langle\omega\rangle$, was near the maximum of the kaon
photoproduction cross-section at 1--1.2\,GeV, which excites a total
energy in the virtual-photon-nucleon center-of-mass system, $W^2 =
M_{targ}^2 + 2 \omega M_{targ} - Q^2$, of 1.6--1.7\,GeV. The kinematic
conditions are summarised in Table~\ref{tab:kinematics}.  The
cryogenic target in the spectrometer hall that was used for the kaon
electro-production experiments consists of a 49.5\,mm long target cell
and is made of a 10\,$\mu$m Havar walls. The geometry of the cell is
optimized for enhancing the luminosity while keeping low the
energy-losses.

% -------------------------------------------------------------------
\section{Kaon Identification}
% -------------------------------------------------------------------

\begin{figure}
  \centering
  \includegraphics[width=0.49\textwidth]{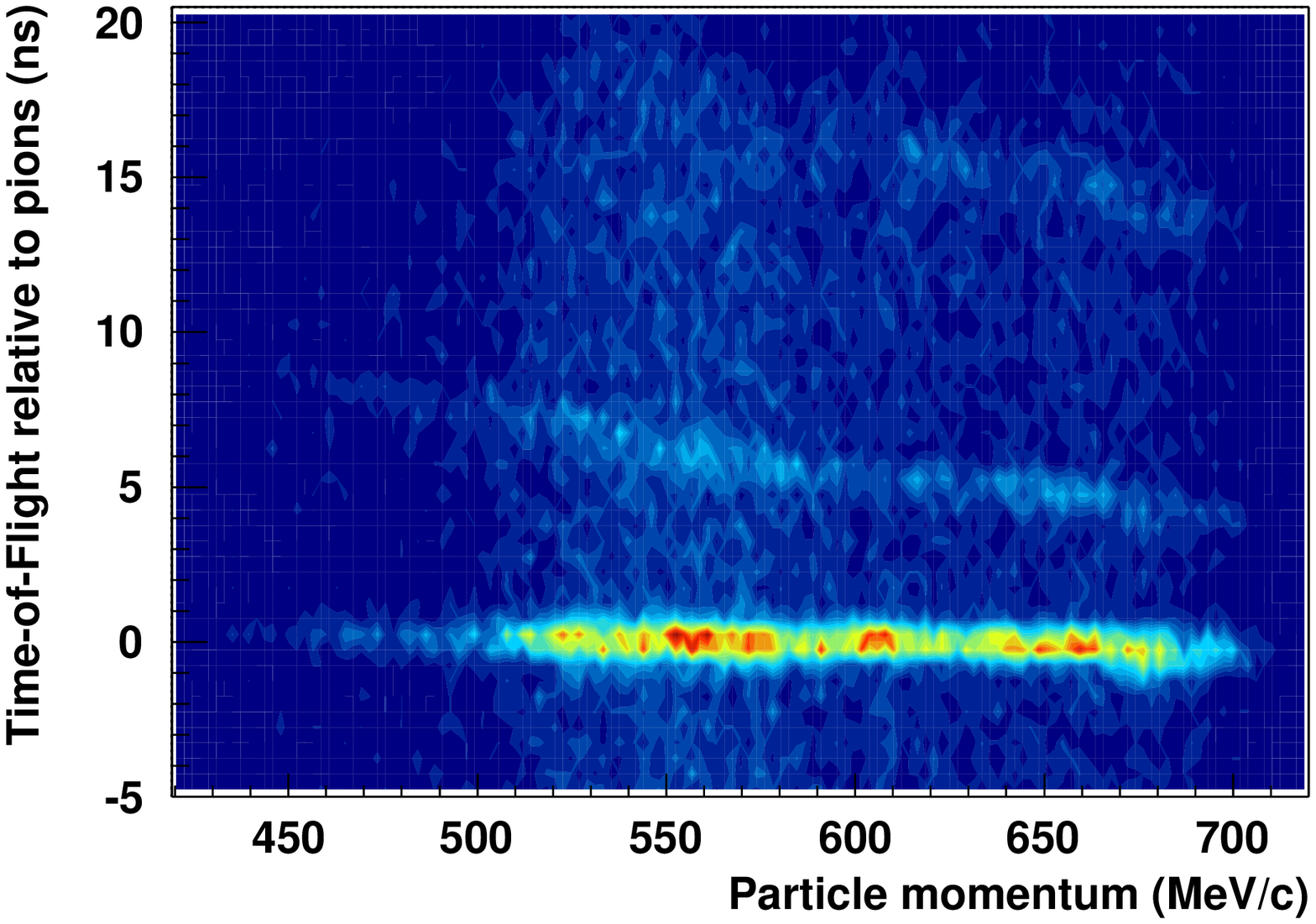}
  \includegraphics[width=0.49\textwidth]{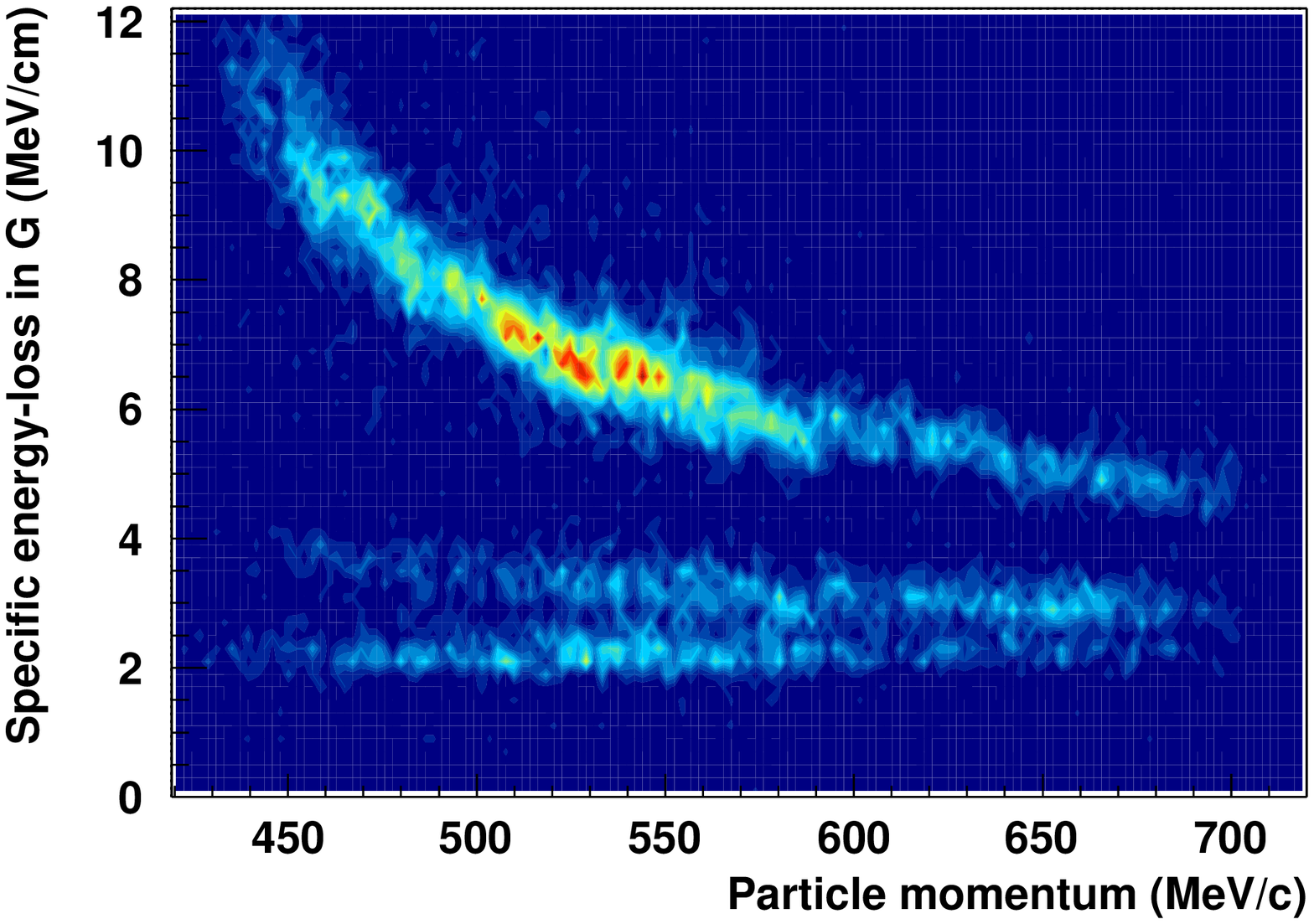}
  \caption{Measured time-of-flight (left) and specific energy-loss
    (right) spectra in scintillator wall~G as a function of particle
    momentum.  The time-of-flight was corrected for the path-length
    dispersion assuming the particle was a pion. The event sample was
    enriched in kaon events by a cut in missing mass. Protons, kaons,
    and pions (curves from top to bottom) are separated.}
  \label{fig:TOFdEdx}
\end{figure}

Kaon identification is based on specific energy-loss and
time-of-flight, electron identification on a signal in a gas Cherenkov
counter.  Measured time-of-flight and specific energy-loss spectra are
shown in Fig.~\ref{fig:TOFdEdx} as a function of particle
momentum. The time-of-flight was corrected for the path-length
dispersion assuming the particle was a pion. The data sample is
dominated by background particles (p and $\pi$) because of a
kaon-to-background ratio of 1\,:\,200 and the event sample was
enriched in kaon events by a cut in missing mass to show the position
of the band of kaons in relation to the two bands from the dominant
particle species.  The flight time difference between protons and
kaons is 10--15\,ns, between pions and kaons 5--10\,ns. The
energy-loss separation between kaons and pions is small, whereas the
separation between kaons and protons is of the order of
2--5\,MeV$\!/$cm.

\begin{figure}
  \includegraphics[width=0.49\textwidth]{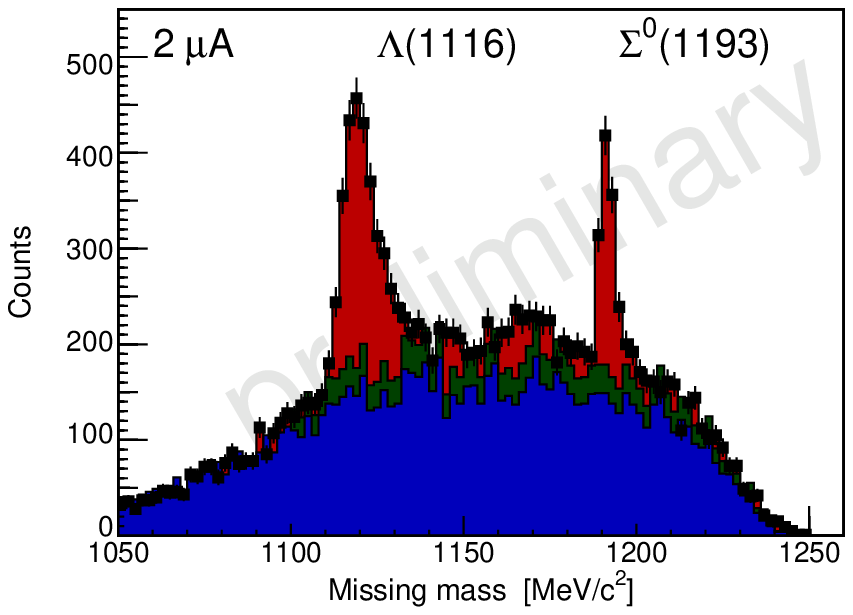}
  \includegraphics[width=0.49\textwidth]{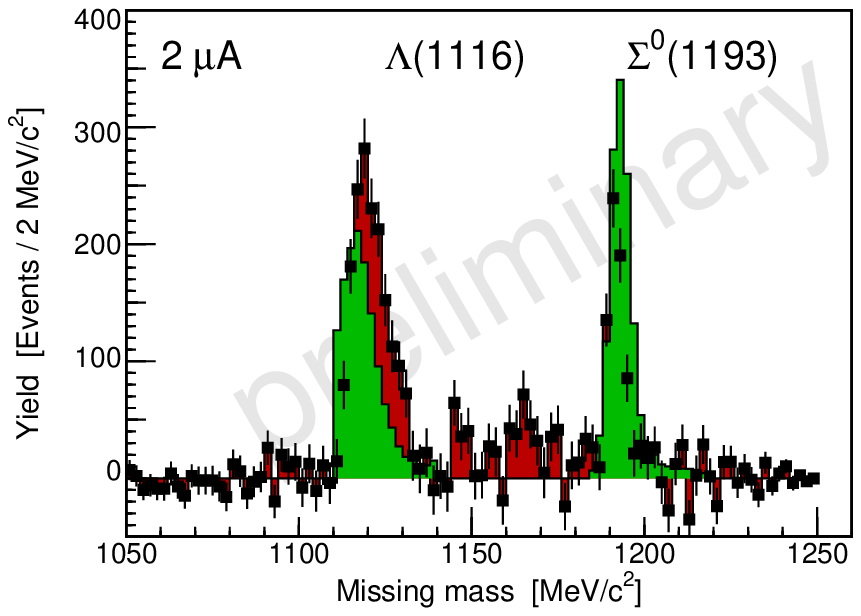}
  \caption{Preliminary missing mass spectra in the p$(e,e'K^+)Y$
    reaction. Left: Missing mass distribution with overlaid histograms
    showing random coincidences in two averaged $(e',K)$
    coincidence-time side-bands with the appropriate weights and
    additional kaon background. The $\Lambda$ and $\Sigma^0$ hyperon
    peaks are evident. Right: Background-subtracted missing mass
    distribution compared to the Monte Carlo spectrum. The simulation
    includes radiative corrections, energy-loss in the target, and a
    model for the resolutions.}
  \label{fig:MissingMass}
\end{figure}

After electron and kaon identification the measured momenta, in
magnitude and direction with respect to the incoming beam, allow for a
full reconstruction of the missing energy and missing momentum of the
recoiling system.  The missing mass $M_X$ is related to the
four-momentum $q^\mu$ of the virtual photon and the four-momentum
$p_K^\mu$ of the detected kaon according to $ M^2_X = (q^\mu +
P^\mu_{targ} - p^\mu_K)^2$, where $P^\mu_{targ} = (M_{targ}, \vec{0})$
is the target four-momentum. The missing energy and missing momentum
can then be calculated using
\begin{eqnarray}
  E_X & = & E_e - E_{e'} + M_{targ} - E_K = \omega + M_{targ} - E_K,\\
  \vec{P}_X & = & \vec{q} - \vec{p}_K,
\end{eqnarray}
and the missing mass in terms of the kinematic variables $Q^2$,
$\omega$, and $\vec{q}$ as well as of the reconstructed kaon energy,
momentum, and scattering angle, can be written as
\begin{equation}
  M_X = \sqrt{E^2_X - |\vec{P}_X|^2} = \sqrt{(\omega - E_K +
    M_{targ})^2 + p_K^2 - 2 p_K |\vec{q}| \cos{\theta_K} - Q^2}\,.
\end{equation}
A preliminary missing mass spectrum is shown in
Fig.~\ref{fig:MissingMass}\,(left). The overlaid histograms show the
missing mass distributions in two averaged $(e',K)$ coincidence-time
side-bands with the appropriate weights and additional kaon
background.  The large momentum acceptance of the \kaos\ spectrometer
covers the free electro-produced hyperons $\Lambda$ and $\Sigma^0$.
The well-known masses of the two hyperons can serve as an absolute
mass calibration. Fig.~\ref{fig:MissingMass}\,(right) shows the
background-subtracted missing mass distribution compared to the Monte
Carlo spectrum. The simulation includes radiative corrections,
energy-loss in the target, and a model for the resolutions.

%

% -------------------------------------------------------------------
\section{Reaction Yields}
% -------------------------------------------------------------------

\begin{figure}
  \includegraphics[width=0.6\textwidth]{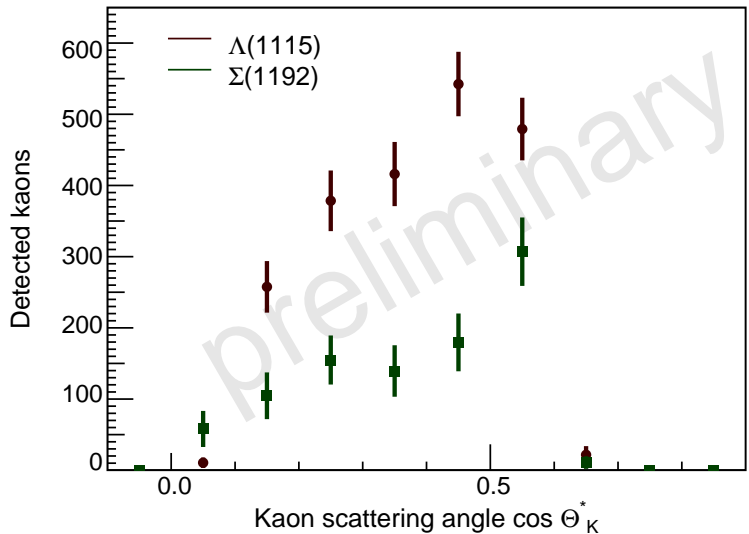}
  \caption{Preliminary yields of identified
    p$(e,e'K^+)\Lambda,\Sigma^0$ events as a function of the cosine of
    the kaon centre-of-mass scattering angle. The identified kaon
    counts in both hyperon channels were corrected for background
    counts to obtain the yields. The 2009 data covered angles in the
    range of $\theta^*_K \sim$ 40--90$^\circ$ and $\phi\sim$
    -20--20$^\circ$. In 2011 scattering angles down to zero degree are
    available for forward-angle measurements of the process.}
  \label{fig:KYield}
\end{figure}

The kaon events that correspond in the missing mass to the two hyperon
channels are used to extract the kaon yield. The preliminary yields of
identified p$(e,e'K^+)\Lambda,\Sigma^0$ events as a function of the
centre-of-mass kaon scattering angle are shown in
Fig.~\ref{fig:KYield}.

The further analysis required a detailed Monte Carlo simulation of the
experiment in order to extract cross-section information from the
data.  For an unpolarised electron beam and an unpolarised target, the
five-fold differential cross-section for the p$(e,e'K^+)\Lambda$
process can be written, see {\em e.g.\/}~\cite{Amaldi1979}, in a very
intuitive form:
\begin{equation}
  \frac{\D\sigma}{\D E_{e'} \D\Omega_{e} \D\Omega_{K}^{*}} = \Gamma(Q^2,W) 
  \frac{\D\sigma_v}{\D\Omega_K^*}(W, Q^2, \epsilon, \theta_K^*, \phi)
\end{equation}
where the virtual photo-production cross-section is conventionally
expressed as
\begin{equation}
   \frac{\D\sigma_v}{\D\Omega_K^*}
   = \frac{\D\sigma_T}{\D\Omega_K^*}
   + \epsilon \frac{\D\sigma_L}{\D\Omega_K^*}
   + \sqrt{2 \epsilon (1 + \epsilon)} \,
     \frac{\D \sigma_{LT}}{\D\Omega_K^*} \cos\phi
   + \epsilon \frac{\D \sigma_{TT}}{\D\Omega_K^*} \cos 2 \phi\, .
\label{eq:diffxsec}
\end{equation}
The kaon polar angle $\theta_K^*$ is given in spherical coordinates in
the hadronic centre-of-mass system. In this system $\omega^*$ is the
energy of the virtual photon. The terms indexed by $T, L, LT, TT$ are
the transverse, longitudinal and interference cross-sections. The
transformation between the differentials $\D E_{e'} \D\Omega_e
\leftrightarrow \D Q^2 \D W $ is incorporated into the virtual photon
flux $\Gamma(Q^2,W)$ via the Jacobian:
\begin{equation}
  \frac{\D\sigma}{\D Q^{2}\D W \D\phi_{e} \D\Omega_{K}^{*}} =
  \frac{W}{2M_{p} EE'} \Gamma(E',\Omega_e)
  \frac{\D\sigma_v}{\D\Omega_K^*} = \Gamma(Q^2,W) \frac{\D\sigma_v}{\D
    \Omega_K^*}
\end{equation}
The experimental yield can then be related to the cross-section by
\begin{equation}
  Y= {\cal L} \times \int\!\! \Gamma(Q^2,W) \frac{\D\sigma_v}{\D
    \Omega_K^{*}} A(\D V) R(\D V)\, \D Q^{2} \D W \D \phi_{e} \D
  \Omega_{K}^{*}\,,
\end{equation}
where ${\cal L}$ is the experimental luminosity that included global
efficiencies such as dead-times and beam-current dependent corrections
such as the tracking efficiency, $A$ is the acceptance function of the
coincidence spectrometer set-up, and $R$ is the correction due to
radiative or energy losses.

With a run-time of 244\,h using 2\,$\mu$A beam current (at 14\,\%
dead-time) and 38\,h using 4\,$\mu$A beam current (at 45\,\%
dead-time) the accumulated and corrected luminosity for the 2009
data-taking campaign was $\int\! {\cal L}\D t \sim$
2300\,fbarn$^{-1}$.
 
The geometrical acceptance of the \kaos\ spectrometer set-up, the
path-length from target to detectors, kaon decay in flight, and kaon
scattering were determined by a Monte Carlo simulation using the
simulation package {\sf Geant4}.  Within the A1 collaboration a
different simulation package ({\tt Simul++}) for the experiments was
developed in the past. This code allows to simulate, according to a
chosen kinematics, the     phase-space accepted by the
spectrometers. In alternative, the events can be generated sampling a
distribution given by the cross-section of a specific process.  The
output of the {\sf Geant4} study was implemented in this simulation as
a generalised acceptance map, leading to the phase-space accepted by
the combination of the kaon spectrometer with the electron
spectrometer together with the radiative corrections.  By virtue of
the Monte Carlo technique used to perform this integral the quantities
$A$ and $R$ are not available separately or on an event-by-event
basis.

The general acceptance can be written in terms of the cross-section at
a given point $({d^{2}\sigma}/{d\Omega_K^*})_{0,\theta_K^*}$, called
the scaling point, at $\langle Q^2 \rangle$, $\langle W \rangle$,
$\langle \phi_e \rangle$, and $\langle \phi \rangle$, when
studying the dependence on the remaining variable $\cos\theta_K^*$:
\begin{eqnarray}
  Y & = & {\cal L} \times \left( \frac{\D\sigma_v}{\D \Omega_K^*}
  \right)_{0,\theta_K^*} \times \\ & & \int\!\! \Gamma(Q^2,W)
  \frac{{\D\sigma_v}/{\D \Omega_K^*}}{ \left( {\D\sigma_v}/{\D
      \Omega_K^*} \right)_{0,\theta_K^*}} A(\D V) R(\D V)\, \D Q^{2}
  \D W \D \phi_{e} \D \Omega_{K}^{*} \nonumber
\end{eqnarray}

Then, the behaviour of the cross-section across the acceptance needs
to be scaled according to a given theoretical description. The
Kaon-Maid~\cite{MAID,Mart1999} isobar model was implemented in this
analysis, however, one can as well use the full (SL) and the
simplified version (SLA) of the so-called Saclay-Lyon
model~\cite{Mizutani1998}. The scaling point for the data set from
2009 is at $\langle Q^2 \rangle =$ 0.036\,(GeV$\!/c)^2$, $\langle W
\rangle =$ 1.750\,GeV, $\langle \epsilon \rangle= $ 0.4 and
$\langle \phi \rangle =$ 0.

% -------------------------------------------------------------------
\section{Cross-Sections}
% -------------------------------------------------------------------

\begin{figure}
  \includegraphics[width=0.49\textwidth]{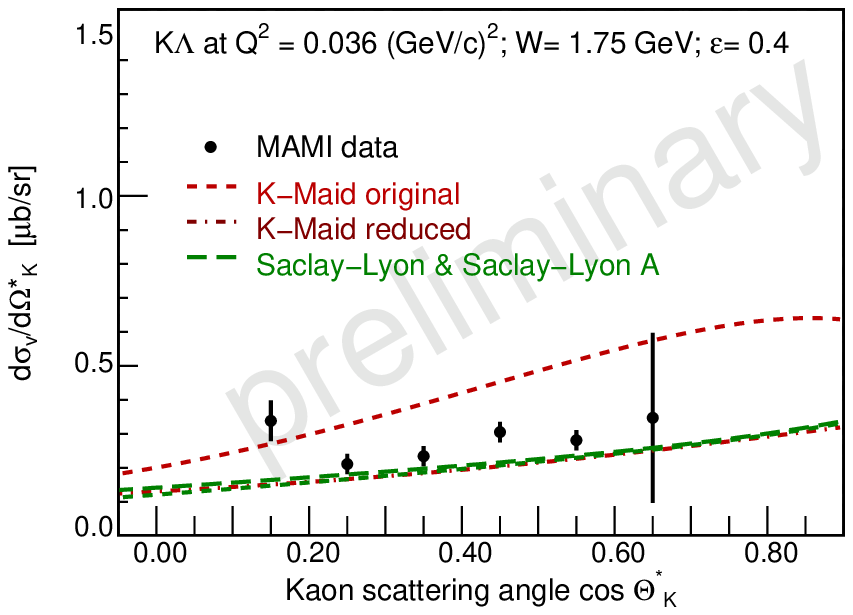}
  \includegraphics[width=0.49\textwidth]{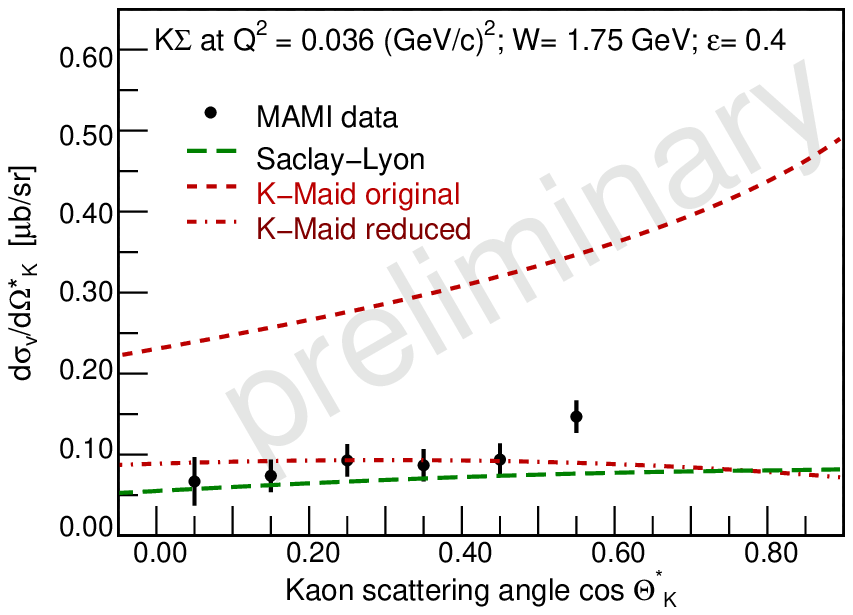}
  \caption{Preliminary differential cross-sections of kaon
    electro-production scaled to the centre of the experimental
    acceptance and compared to variants of the Kaon-Maid model and of
    the Saclay-Lyon model.}
  \label{fig:CrossSections}
\end{figure}

Finally, the cross-section is extracted dividing the yield by the
luminosity and the evaluated integral. The angular distribution of the
cross-section gives information on the angular dependence of the
production mechanisms for each hyperon.  Preliminary differential
cross-sections of kaon electro-production in the $\Lambda$ and
$\Sigma^0$ channels are shown in Fig.~\ref{fig:CrossSections}.  The
data points are compared to the Kaon-Maid model, to a variant of that
model, and to the Saclay-Lyon model. It is expected that the final
results will constrain the existing phenomenological models and
generate theoretical interest.  The difference between photo- and
electroproduction data at small $Q^2$ is very important as various
models predict different transitions between the photoproduction point
and the electroproduction cross-section.  If, {\em e.g.}, the
Kaon-MAID model was right, then the electroproduction data would be at
$\cos\theta^*_K=$ 0.5 a factor 2.5 larger than the photoproduction
data. The data then could help to fix the unknown longitudinal
coupling constants in this kinematical region.

%

% -------------------------------------------------------------------
\section{Perspectives}
% -------------------------------------------------------------------

\begin{figure}
  \includegraphics[width=0.6\textwidth]{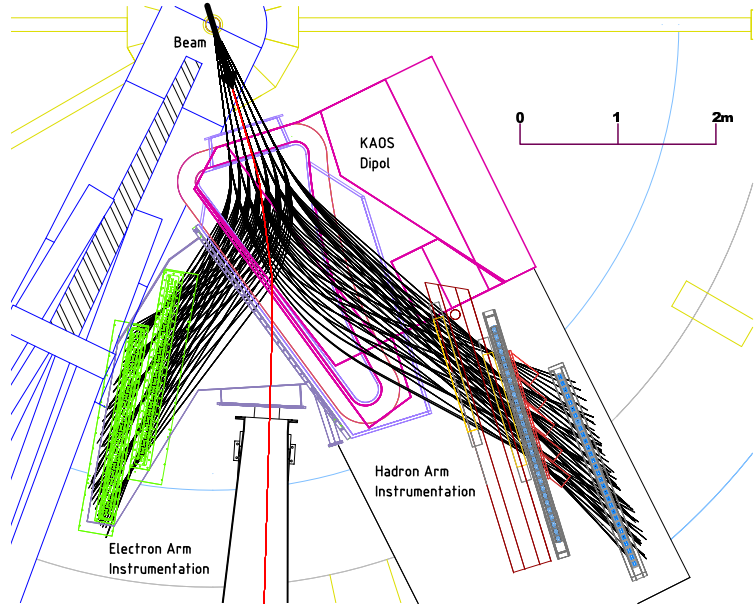}
  \caption{Layout of the coming experiments with the
    \kaos\ spectrometer at zero degree scattering angle. Charged
    particles will be detected on both sides of the beam exit,
    ray-traced trajectories are shown by full lines.  The
    instrumentation of the two spectrometer arms is indicated.}
  \label{fig:Kaos}
\end{figure}

Using the recently installed magnetic spectrometer \kaos\ and a
high-resolution spectrometer in coincidence it was possible measure
elementary kaon electro-production at MAMI in a kinematic regime not
covered by other experiments. The most interesting kinematic region is
at at very forward laboratory angles.  Precise data from very forward
angle kaon production are urgently required for calculating
hypernuclear production cross-sections. In this kinematics the
elementary amplitude serves as the basic input, which determines the
accuracy of predictions for hypernuclei~\cite{Bydzovsky2006}.  The
detection of kaons in this angular range and the tagging of nearly
zero-degree electrons will be achieved at MAMI in the very near future
by steering the primary beam through the \kaos\ spectrometer after
passing through a magnetic chicane comprising two compensating sector
magnets.  The compact design of \kaos\ and its capability to detect
negatively and positively charged particles simultaneously under
forward scattering angles complement the existing
spectrometers. Fig.~\ref{fig:Kaos} shows the layout of the coming
experiments with the \kaos\ spectrometer at zero degree.  In order to
use \kaos\ as a zero-degree double-arm spectrometer a new electron-arm
focal-plane detector system was built. It consists of two vertical
planes of 18,432 fibres. Detectors and electronics for the 4,608
read-out and level-1 trigger channels have been installed and tested.

For the year 2011 elementary kaon electroproduction measurements with
the \kaos\ spectrometer at zero degree scattering angle using the
commissioned beam chicane and a first hypernuclear decay-pion
spectroscopy experiment using the \kaos\ spectrometer as kaon tagger
are planned. A feasibility study on the decay-pion spectroscopy in a
single-arm measurement was performed in 2010.

\begin{acknowledgments}

We acknowledge the generous help that we have received from the
accelerator group of MAMI.

We acknowledge our gratitude support from the Federal State of
Rhineland-Palatinate and by the Deutsche Forschungsgemeinschaft with
the Collaborative Research Center 443.

\end{acknowledgments}

%\bibliographystyle{epj}
%\bibliography{../BIBLIOGRAPHY/literature-2010}

\end{document}